\documentclass[iop]{emulateapj}
\pdfoutput=1
\usepackage{color}
\usepackage{amsmath}
\usepackage{mathtools}
\usepackage{multirow}
\usepackage{subfigure}

\newcommand{\dnds}{$\mathrm{d}N/\mathrm{d}S$}

\newcommand{\pzc}{\mbox{Z16}}
\newcommand{\opdf}{1pPDF}
\newcommand{\shy}{\mathrm{h}}

\slugcomment{Manuscript accepted for publication in ApJL (2016 July 1)}

\shortauthors{Zechlin et al.}

\begin{document}

\title{STATISTICAL MEASUREMENT OF THE GAMMA-RAY SOURCE-COUNT DISTRIBUTION AS A FUNCTION OF ENERGY}

\author{Hannes-S. Zechlin\altaffilmark{1,2}, Alessandro Cuoco\altaffilmark{2,3},\\
        Fiorenza Donato\altaffilmark{1,2}, Nicolao Fornengo\altaffilmark{1,2}, and
        Marco Regis\altaffilmark{1,2}}

\altaffiltext{1}{Dipartimento di Fisica, Universit\`a di Torino, via P. Giuria, 1, I-10125 Torino, Italy; zechlin@to.infn.it}
\altaffiltext{2}{Istituto Nazionale di Fisica Nucleare, Sezione di Torino, via P. Giuria, 1, I-10125 Torino, Italy}
\altaffiltext{3}{Institute for Theoretical Particle Physics and Cosmology (TTK), RWTH Aachen University, D-52056 Aachen, Germany}

\begin{abstract}
Statistical properties of photon count maps have recently been proven
as a new tool to study the composition of the gamma-ray sky with high
precision. We employ the 1-point probability distribution function of
6 years of \textit{Fermi}-LAT data to measure the source-count
distribution \dnds\ and the diffuse components of the high-latitude
gamma-ray sky as a function of energy.  To that aim, we analyze the
gamma-ray emission in five adjacent energy bands between 1\,GeV and
171\,GeV. It is demonstrated that the source-count distribution as a
function of flux is compatible with a broken power law up to energies
of \mbox{$\sim\! 50\,\mathrm{GeV}$}. The index below the break is
between 1.95 and 2.0\,. For higher energies, a simple power-law fits
the data, with an index of $2.2^{+0.7}_{-0.3}$ in the energy band
between 50\,GeV and 171\,GeV.  Upper limits on further possible breaks
as well as the angular power of unresolved sources are derived. We
find that point-source populations probed by this method can explain
$83^{+7}_{-13}$\% ($81^{+52}_{-19}$\%) of the extragalactic gamma-ray
background between 1.04\,GeV and 1.99\,GeV (50\,GeV and 171\,GeV).
The method has excellent capabilities for constraining the gamma-ray
luminosity function and the spectra of unresolved blazars.
\end{abstract}

\section{INTRODUCTION}\label{sec:intro}
The study of gamma-ray sources is one of the main tasks in
understanding the gamma-ray sky.  Compact emission regions and high
distances cause most extragalactic sources to appear point-like in
current measurements \citep[][]{2015ApJS..218...23A}, in distinction
to truly diffuse components such as Galactic foreground (GF) emission
\citep[][]{2016arXiv160207246A} or unresolved components like the
isotropic diffuse gamma-ray background
\citep[IGRB;][]{2015ApJ...799...86A}. Source populations are
characterized by individual source-count distributions, which encode
their physical and evolutional properties as functions of the integral
flux $S$. Since different source classes, e.g., flat-spectrum radio
quasars (FSRQs), BL Lacertae (BL Lac) objects, misaligned active
galactic nuclei (mAGN), and star-forming galaxies (SFGs) distinguish
themselves by spectral index, distance, and luminosity
\citep[e.g.,][]{2016A&ARv..24....2M}, the combined gamma-ray
source-count distribution \dnds\ is nontrivial, and depends on the
energy band considered.

For the energy band between 1\,GeV and 10\,GeV,
\citet{2015arXiv151207190Z}, henceforth \pzc, have recently
demonstrated that statistical methods can be used to measure the
combined \dnds\ with high accuracy and with sensitivity down to fluxes
about one order of magnitude below catalog detection thresholds.  The
6-year data provided by the \emph{Fermi} Large Area Telescope
\citep[\emph{Fermi}-LAT;][]{2009ApJ...697.1071A,2012ApJS..203....4A}
have been used to determine \dnds\ and the composition of the
gamma-ray sky at high Galactic latitudes ($|b|\geq 30^\circ$). The
analysis was based on a global fit of the 1-point probability
distribution function (1pPDF) of the LAT photon counts map. The method
of analyzing the simple 1pPDF \citep[cf.][]{2011ApJ...738..181M} has
been extended to include spatial templates, providing significantly
higher sensitivity.  Furthermore, the 1pPDF method does not introduce
biases in the measured \dnds\ distribution near the catalog detection
threshold.  A similar analysis has been recently applied to the
Galactic Center region \citep{2016PhRvL.116e1103L}.

In this Letter, we extend the analysis of \pzc\ to five energy bands,
covering the range from 1.04\,GeV to 171\,GeV, thus measuring
\dnds\ as a function of energy.  We use the same data and methods
described in \pzc.  The measurements are compared to the
\dnds\ distribution predicted by fiducial models of extragalactic
source populations and implications are discussed.

\vspace{-0.3cm}

\section{METHOD AND DATA}\label{sec:method}
The gamma-ray sky was modeled with a superposition of three
components: (i) an isotropic distribution of point sources \dnds, (ii)
a contribution from GF emission, and (iii) a truly isotropic
component, representing sources too faint to be seen by the
\opdf\ method or residual cosmic-ray contamination. For each energy
band, the \dnds\ distribution was approximated with a multiply broken
power law (MBPL) with $N_\mathrm{b}$ breaks $S_{\mathrm{b}j}$,
$j=1,2,\dots,N_\mathrm{b}$,
\vspace{-0.3cm}
\begin{equation}\label{eq:mbpl}
\frac{\mathrm{d}N}{\mathrm{d}S} \propto
\begin{cases} 
\left( \frac{S}{S_0} \right)^{-n_1} \qquad\qquad\qquad\qquad\qquad ,
S > S_{\mathrm{b}1} & \\
\left( \frac{S_{\mathrm{b}1}}{S_0} \right)^{-n_1+n_2}
\left( \frac{S}{S_0} \right)^{-n_2}  \qquad\qquad , 
S_{\mathrm{b}2} < S \leq S_{\mathrm{b}1} \\
\vdotswithin{\left(\frac{S_{\mathrm{b}1}}{S_0}\right)}
\qquad\qquad\qquad\qquad\qquad \vdotswithin{S_{\mathrm{b}2} < S} & \\
\left( \frac{S_{\mathrm{b}1}}{S_0} \right)^{-n_1+n_2}
\left( \frac{S_{\mathrm{b}2}}{S_0} \right)^{-n_2+n_3}
\cdots \ \left( \frac{S}{S_0} \right)^{-n_{N_\mathrm{b}+1}} & \\
\hspace{14em} , S \leq S_{\mathrm{b}N_\mathrm{b}}, & \\
\end{cases}
\end{equation}
where $S_0$ is a constant reference flux and the indices of the
power-law components are denoted by $n_j$.  For simplicity, the
notation omits a separate index denoting the energy dependence of
\dnds.  The \opdf\ was computed following the method of \pzc.
Exposure inhomogeneities were accounted for by slicing the map into
$N_\mathrm{exp} = 20$ regions separated by iso-contours of equal
exposure. The GF and the modeling of the diffuse isotropic background
component were treated in the same way as in \pzc.\footnote{Masking
  Galactic structures such as the Fermi~Bubbles or Galactic~Loop~I did
  not significantly affect the results.}  Corrections of \dnds\ for
the finite point-spread function (PSF) of the instrument were computed
consistently for each energy band considered in this analysis. The
data were fit as described in \pzc, adopting the definition of the
likelihood function $\mathcal{L}_2 ({\bf \Theta})$ (see \pzc,
Equation~17). Besides the normalization of \dnds, the vector of free
parameters ${\bf \Theta}$ contains the breaks $S_{\mathrm{b}j}$, the
indices $n_j$, an overall normalization $A_\mathrm{gal}$ of the GF
template, and the (integral) flux $F_\mathrm{iso}$ of the diffuse
isotropic background component.  The likelihood was sampled using
\texttt{MultiNest} \citep[][]{2009MNRAS.398.1601F}. We used the
frequentist parameter estimation approach, based on the profile
likelihood function as derived from the likelihood samples obtained.

We considered \emph{Fermi}-LAT data (\texttt{P7REP\_CLEAN}) covering
the first 6~years of the mission science operations for five adjacent
energy bands. Table~\ref{tab:ebins} lists the energy bands and
corresponding analysis parameters.

\begin{deluxetable}{lccccccc}
\tablecaption{Energy Bands and Analysis Parameters\label{tab:ebins}}
\tablewidth{0pt}
\tablehead{
\colhead{$E_\mathrm{min}$} & \colhead{$E_\mathrm{max}$} & \colhead{$|b|$} & \colhead{$\kappa$} 
& \colhead{$\sigma_\mathrm{psf}$} & \colhead{$\Gamma$} & \colhead{$S_0/10^{-9}$} & \colhead{$N^\shy_\mathrm{b}$} \\
\colhead{$(\mathrm{GeV})$} & \colhead{$(\mathrm{GeV})$} & \colhead{$(^\circ)$} &   & \colhead{$(^\circ)$} &   &
\colhead{$(\mathrm{cm}^{-2}\mathrm{s}^{-1})$} &  
}
\startdata
$1.04$ & $1.99$ & $\geq 30$ & $6$ & $0.52$ & $2.4$ & $30$ & 1,\,2,\,3 \\
$1.99$ & $5.0$ & $\geq 30$ & $6$ & $0.31$ & $2.4$ & $5$ & 1,\,2,\,3 \\
$5.0$ & $10.4$ & $\geq 30$ & $6$ & $0.23$ & $2.4$ & $1$ & 1,\,2 \\
$10.4$ & $50.0$ & $\geq 30$ & $6$ & $0.15$ & $2.2$ & $0.1$ & 1,\,2 \\
$50$ & $171$ & $\geq 10$ & $7$ & $0.13$ & $2.2$ & $0.1$ & 1,\,2 
\enddata
\tablecomments{For each energy band $[E_\mathrm{min},E_\mathrm{max}]$,
  the table lists the considered Galactic latitude cut $|b|$, the
  resolution of the HEALPix pixelization $\kappa$, where the number of
  pixels of an all-sky map is given by \mbox{$12\times 2^{2\kappa}$},
  and the 68\% containment radius of the effective PSF,
  $\sigma_\mathrm{psf}$. Point sources were assumed to emit
  power-law-type energy spectra with photon index $\Gamma$.  The
  reference flux chosen for the MBPL parameterization is given by
  $S_0$.  The last column lists the numbers of free breaks considered
  for the hybrid approach.}
\end{deluxetable}

The choice of the energy bands was motivated by the analysis of
gamma-ray anisotropies in the high-latitude sky; see
\citet{Ackermann:2012uf}. The highest-energy band matches the first
energy bin quoted in the 2FHL catalog \citep{2016ApJS..222....5A}.
Event selection and data processing were carried out as outlined in
\pzc\ for every individual energy band. We allowed a maximum zenith
angle of $90^\circ$ and the rocking angle of the satellite was
constrained to values smaller than $52^\circ$. To avoid unnecessary
broadening of the effective PSF, we restricted the event selection to
\texttt{FRONT}-converting events for the two lowest-energy bands,
while for higher energies all events were used.  The resolution
parameter $\kappa$ of the HEALPix pixelization
\citep{2005ApJ...622..759G} was chosen to undersample the effective
PSF (see Section~3 in \pzc). We compared the resolutions $\kappa=6$
and $\kappa=7$, adopting the one providing higher sensitivity.  Source
spectral energy distributions were assumed to follow power laws.  The
average spectral photon index $\Gamma$ was selected following
\citet{2010ApJ...720..435A}.\footnote{We checked that systematic
  uncertainties related to this choice are small, by varying $\Gamma$
  between 2.1 and 2.4\,.}  For all but the last energy band we
restricted the analysis to Galactic latitudes $|b| \geq 30^\circ$. Due
to significantly fewer events, we chose a Galactic latitude cut of
$|b| \geq 10^\circ$ for the highest-energy band.  Indeed, in this
band, the GF is less prominent, its spectrum being softer than the
source component.

\subsection{Source-count Distribution Fit}\label{ssec:dnds_fit}
To fit our model of the gamma sky to the data, we used the analysis
chain developed by \pzc, i.e., \dnds\ was parameterized with a pure
MBPL (MBPL approach) and an improved representation incorporating an
additional node (hybrid approach).  A node is defined as a break at a
fixed position, chosen at the faint end of \dnds, with the
normalization left free to vary.  The reference flux $S_0$ of the MBPL
representation of \dnds\ was chosen for each energy band separately
(see Table~\ref{tab:ebins}). Stability was checked by varying $S_0$
within a factor of 10.  We compared MBPL parameterizations of the
\dnds\ distribution with up to three free breaks, depending on the
energy band.

\section{RESULTS AND DISCUSSION}\label{sec:results}
\subsection{Source-count Distribution}\label{ssec:res_dnds}
The MBPL approach was employed uniformly for each energy band,
comparing \dnds\ parameterizations with two and three free breaks.
The data were described sufficiently by two-break fits for all energy
bands, i.e., no statistical preference for three breaks was found,
with the test statistic $\mathrm{TS} = -2\,\Delta \ln \mathcal{L}_2$
between the two hypotheses at most reaching a value of 2. As discussed
in \pzc, below the second (third) break the fit generally prefers a
sharp cutoff, which we interpret as a loss of sensitivity of the
method (as opposed to an intrinsic feature of the
\dnds\ distribution).

\begin{figure*}[t]
\begin{centering}
\subfigure[]{%
  \includegraphics[width=0.95\textwidth]{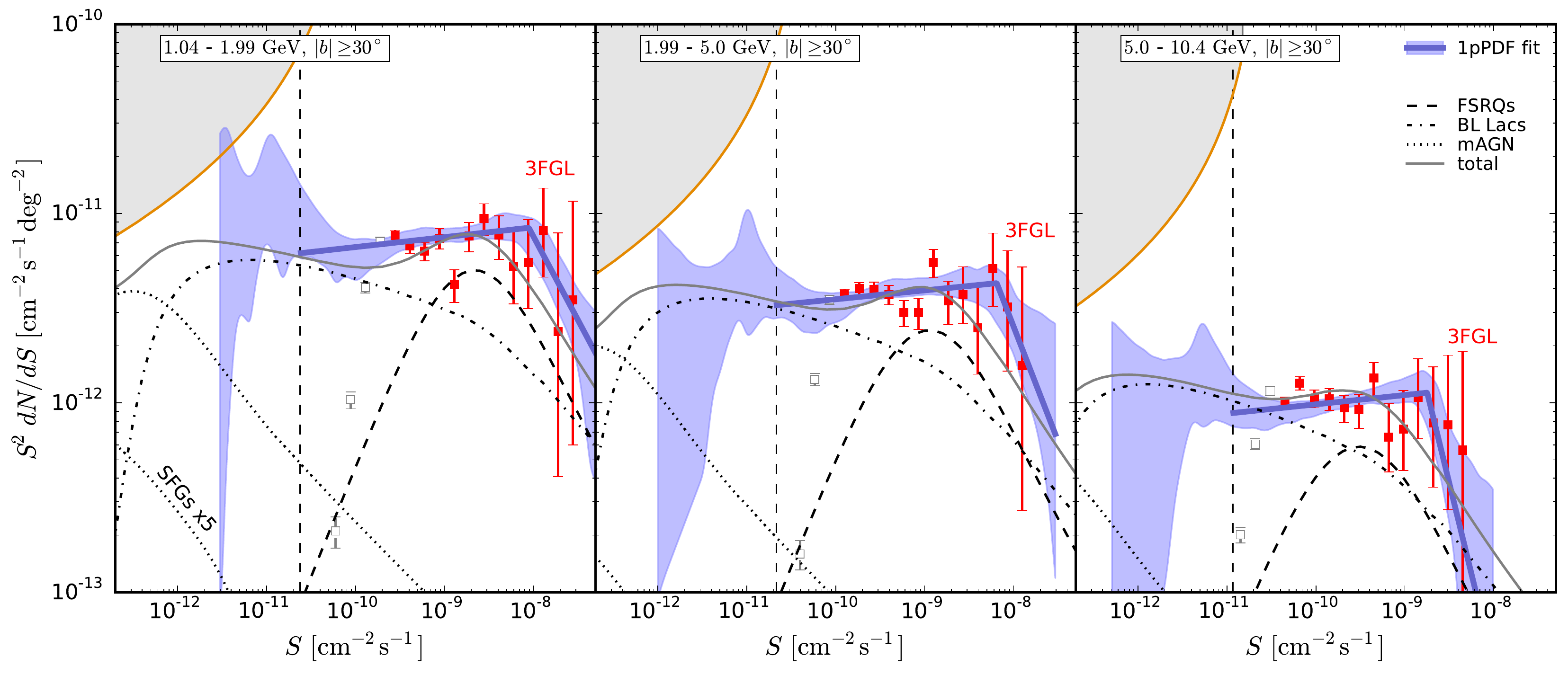}
  \label{sfig:dNdStop}
}\\
\subfigure[]{%
  \includegraphics[width=0.65\textwidth]{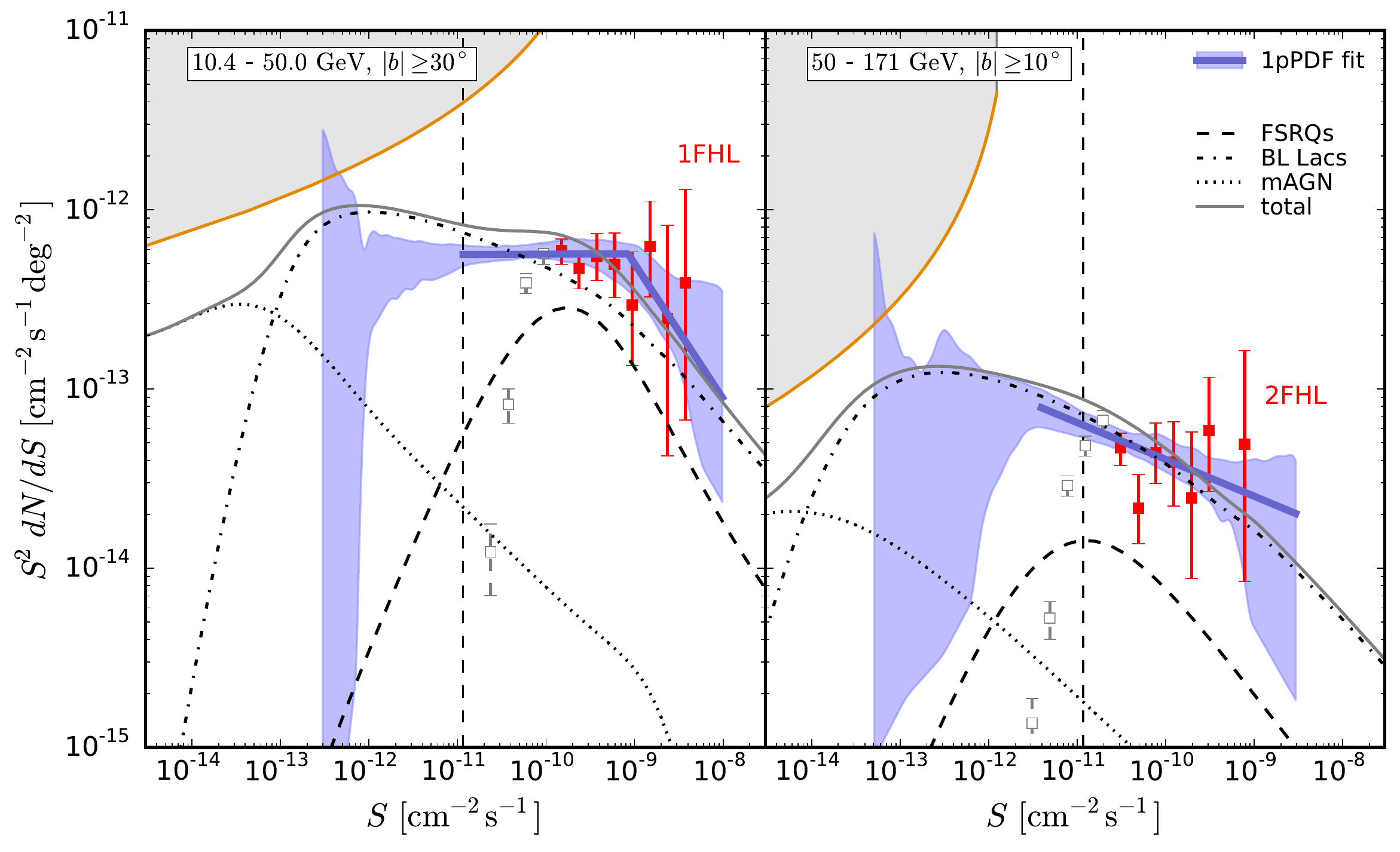}
  \label{sfig:dNdsbottom}
}
\caption{Differential source-count distributions \dnds\ obtained from
  6-year \textit{Fermi}-LAT data with the \opdf\ method. The data have
  been fit in five energy bands using the hybrid approach.  The best
  fits and the uncertainties at 68\% confidence level are depicted by
  the solid blue lines and the blue shaded bands. The fits are
  compared to the \dnds\ distributions derived from cataloged sources
  \citep[red squares; Poissonian errors
    following][]{1986ApJ...303..336G}.  The open gray squares depict
  \dnds\ points from sources below the nominal detection threshold,
  where the sample suffers from catalog incompleteness. Thus, they
  cannot be used for comparison.  The vertical dashed lines indicate
  the sensitivity estimates $S_\mathrm{sens}$. The orange lines,
  bordering the gray shaded areas, mark the region where the
  contribution from point sources equals $F_\mathrm{ps} =
  F_\mathrm{tot} - F_\mathrm{gal}$.  In this case, $F_\mathrm{ps}$ has
  been derived with Equation~\eqref{eq:F_ps}, but extrapolating the
  best-fit \dnds\ distributions with a power-law component of varying
  index below the position of the last free break. The orange lines
  therefore constrain the position of a next break, given the
  condition $F_\mathrm{ps} \leq F_\mathrm{tot} - F_\mathrm{gal}$.  The
  dashed, dot-dashed, and dotted lines depict model predictions for
  FSRQs \citep{2012ApJ...751..108A}, BL~Lacs \citep[][model
    LDDE$_1$]{2014ApJ...780...73A}, and mAGN
  \citep{2014ApJ...780..161D}, respectively. The model of SFGs has
  been taken from \citet{2012ApJ...755..164A}. The solid gray lines
  denote the sum of these contributions. (The data used to create this
  figure are available.)\label{fig:dNdS}}
\end{centering}
\end{figure*}

To improve the estimation of uncertainty bands, we employed the hybrid
approach by adding a node $S_\mathrm{nd}$, imposing the faint cutoff
positions found with the MBPL approach.  We chose nodes at $3\times
10^{-12}$, $10^{-12}$, $5\times 10^{-13}$, $3\times 10^{-13}$, and
$5\times 10^{-14}\,\mathrm{cm}^{-2}\,\mathrm{s}^{-1}$ for the bands as
ordered by increasing energy.  The power-law index below the node was
fixed to $-10$.  Given the high number of events detected in the first
two energy bands\footnote{Total number of selected events, ordered by
  energy: 487,854, 268,261, 119,123, 57,571, and 10,005.}, we
considered \dnds\ parameterizations with \emph{one, two, and three}
free breaks (and a node), see Table~\ref{tab:ebins}, as driven by
results of \pzc.  Due to significantly fewer events at higher energies
and the resulting sensitivity decrease, only \emph{one or two} free
breaks were used for the last three energy bands.

\begin{deluxetable*}{lccccc}
\tablecaption{\dnds\ Parameters and Sky Composition\label{tab:dnds_comp}}
\tablewidth{0pt}
\tablehead{
\colhead{Parameter\tablenotemark{a}} & $1.04-1.99\,\mathrm{GeV}$ & $1.99-5.0\,\mathrm{GeV}$ 
& $5.0-10.4\,\mathrm{GeV}$ & $10.4-50.0\,\mathrm{GeV}$ & $50-171\,\mathrm{GeV}$
}
\startdata
$S_\mathrm{b1}$ &  $9^{+10}_{-5} \times 10^{-9}$\phantom{$^0$} &  $7^{+13}_{-5} \times 10^{-9}$\phantom{$^0$}
& $2^{+2}_{-1} \times 10^{-9}$\phantom{$^0$} &  $8^{+21}_{-8}\times 10^{-10}$\phantom{$^0$} &  \nodata \\ 
$n_1$ &  $2.88^{+0.83}_{-0.47}$ &  $3.21^{+1.09}_{-1.16}$ &  $3.92^{+1.72}_{-1.53}$ &  $2.75^{+0.21}_{-0.68}$ &  $2.21^{+0.69}_{-0.29}$ \\
$n_2$ &  $1.95^{+0.06}_{-0.07}$ &  $1.95^{+0.07}_{-0.06}$ &  $1.95^{+0.13}_{-0.09}$ &  $2.00^{+0.21}_{-0.28}$ &  \nodata \\ 
\tableline
$F_\mathrm{ps}$ & $2.5^{+0.2}_{-0.4}\times 10^{-7}$ &  $1.24^{+0.07}_{-0.25}\times 10^{-7}$
&  $2.7^{+0.8}_{-0.3}\times 10^{-8}$ &  $1.4^{+0.6}_{-0.1}\times 10^{-8}$ &  $1.7^{+1.1}_{-0.4} \times 10^{-9}$ \\
$F_\mathrm{gal}$ & $6.54^{+0.02}_{-0.05}\times 10^{-7}$ &  $3.24^{+0.02}_{-0.03}\times 10^{-7}$
&  $6.59^{+0.07}_{-0.10}\times 10^{-8}$ &  $2.60^{+0.04}_{-0.07}\times 10^{-8}$ &  $2.75^{+0.07}_{-0.10} \times 10^{-9}$ \\
$F_\mathrm{iso}$ & $1.8^{+3.9}_{-0.8}\times 10^{-8}$ &  $5.6^{+27.8}_{-0.6}\times 10^{-9}$
& $1.8^{+0.3}_{-0.7}\times 10^{-8}$ &  $1.2^{+0.2}_{-0.8}\times 10^{-8}$ &  $1.2^{+0.5}_{-0.9}\times 10^{-9}$ \\
$F_\mathrm{tot}$ & $9.17(1) \times 10^{-7}$ &  $4.573(9) \times 10^{-7}$
& $1.103(3) \times 10^{-7}$ & $5.27(2) \times 10^{-8}$  &  $5.67(6) \times 10^{-9}$ \\
$q_\mathrm{ps}$ & $0.27^{+0.02}_{-0.04}$ &  $0.27^{+0.02}_{-0.06}$
& $0.24^{+0.08}_{-0.03}$ &  $0.27^{+0.11}_{-0.03}$ &  $0.29^{+0.19}_{-0.08}$ \\
$q_\mathrm{gal}$ & $0.714^{+0.003}_{-0.005}$ &  $0.708^{+0.005}_{-0.006}$
& $0.598^{+0.007}_{-0.01}$ &  $0.494^{+0.008}_{-0.013}$ &  $0.49^{+0.01}_{-0.02}$ \\
$q_\mathrm{iso}$ & $0.02^{+0.04}_{-0.01}$ &  $0.012^{+0.061}_{-0.001}$
& $0.16^{+0.03}_{-0.07}$ &  $0.23^{+0.04}_{-0.15}$ &  $0.22^{+0.10}_{-0.17}$ \\
\tableline
$S^\mathrm{UL}_\mathrm{b1} (\Delta n_{12} > 0.3)$ & \nodata & \nodata
& \nodata & \nodata &  $1.3 \times 10^{-11}$ \\
$S^\mathrm{UL}_\mathrm{b2} (\Delta n_{23} > 0.3)$ & $2.3 \times 10^{-10}$ & $1.7 \times 10^{-10}$
& $1.5 \times 10^{-10}$ & $3.3 \times 10^{-11}$ & \nodata \\
$C_\mathrm{P}(S^\mathrm{3FGL}_\mathrm{th})$ &  $2.3^{+0.7}_{-0.1}\times 10^{-18}$ &  $6.1^{+2.2}_{-0.4}\times 10^{-19}$
&  $5^{+1}_{-1}\times 10^{-20}$ &  $2.3^{+0.4}_{-0.5}\times 10^{-20}$ &  $2^{+3}_{-4}\times 10^{-22}$ \\
$C_\mathrm{P}(S^\mathrm{fix}_\mathrm{th})$ &  $4.2^{+0.7}_{-0.1}\times 10^{-18}$ &  $1.06^{+0.22}_{-0.04}\times 10^{-18}$
&  $1.3^{+0.1}_{-0.1}\times 10^{-19}$ &  $1.8^{+0.1}_{-0.1}\times 10^{-19}$ &  $4.4^{+0.3}_{-0.7}\times 10^{-21}$ 
\enddata
\tablecomments{Selection of parameter values obtained for different
  energy bands.  The parameters of the \dnds\ distribution correspond
  to a parameterization with one break $S_\mathrm{b1}$ and a
  node. Parentheses denote symmetric errors on the preceding
  digit. The quantities $q_\mathrm{ps}$, $q_\mathrm{gal}$, and
  $q_\mathrm{iso}$ are the ratios of the integral flux components and
  $F_\mathrm{tot}$. The upper limits on a first
  ($S^\mathrm{UL}_\mathrm{b1}$) or second
  ($S^\mathrm{UL}_\mathrm{b2}$) intrinsic break are at the 95\%
  confidence level.  The anisotropy $C_\mathrm{P}$ is given for two
  different point-source detection thresholds:
  $C_\mathrm{P}(S^\mathrm{3FGL}_\mathrm{th})$ denotes the anisotropy
  for the effective detection threshold $S^\mathrm{3FGL}_\mathrm{th}$
  of the 3FGL catalog. Since it is difficult to explicitly define
  $S^\mathrm{3FGL}_\mathrm{th}$, the corresponding anisotropy has been
  estimated as $C_\mathrm{P}(S^\mathrm{3FGL}_\mathrm{th}) \approx
  C_\mathrm{P}(S^{<1}_\mathrm{th}) -
  C^\mathrm{cat}_\mathrm{P}(S^{<1}_\mathrm{th})$, where $C_\mathrm{P}$
  refers to Equation~\eqref{eq:Cp}, $C^\mathrm{cat}_\mathrm{P}$
  denotes the anisotropy contributed by cataloged (i.e., resolved)
  sources only, and $S^{<1}_\mathrm{th}$ approximates the flux below
  which the detection efficiency of the 3FGL catalog becomes much less
  than $1$. For $S^{<1}_\mathrm{th}$, the values $2\times 10^{-10}$,
  $10^{-10}$, $4\times 10^{-11}$, $2\times 10^{-11}$, and $8\times
  10^{-12}\,\mathrm{cm}^{-2}\,\mathrm{s}^{-1}$ (from left to right)
  have been used.  The results are stable against choosing higher
  values of $S^{<1}_\mathrm{th}$.  On the contrary,
  $S^\mathrm{fix}_\mathrm{th}$ denotes a sharp threshold,
  approximating the flux above which the catalogs used in
  Figure~\ref{fig:dNdS} have full detection efficiency, i.e., $2\times
  10^{-10}$, $10^{-10}$, $4\times 10^{-11}$, $10^{-10}$, and $2\times
  10^{-11}\,\mathrm{cm}^{-2}\,\mathrm{s}^{-1}$.}
\tablenotetext{a}{The break $S_{\mathrm{b} 1}$ and the upper limits
  $S^\mathrm{UL}_{\mathrm{b} j}$ are given in units of
  $\mathrm{cm}^{-2}\,\mathrm{s}^{-1}$. The integral fluxes
  $F_\mathrm{ps}$, $F_\mathrm{gal}$, $F_\mathrm{iso}$, and
  $F_\mathrm{tot}$ are in units of
  $\mathrm{cm}^{-2}\,\mathrm{s}^{-1}\,\mathrm{sr}^{-1}$. The unit of
  the anisotropy $C_\mathrm{P}$ is
  $(\mathrm{cm}^{-2}\,\mathrm{s}^{-1}\,\mathrm{sr}^{-1})^2\,\mathrm{sr}$.}
\end{deluxetable*}

The results are shown in Figure~\ref{fig:dNdS}. Marginalized parameter
estimates are listed in Table~\ref{tab:dnds_comp}.  In addition, the
best-fit \dnds\ distributions and the corresponding uncertainty bands
are provided as supporting material.

As for the MBPL case, we found that the additional breaks did not
significantly improve the fit for any of the five energy bands.  The
data were described sufficiently well by \dnds\ distributions with a
single break at comparably high fluxes and a node at the faint end.
The best fits for this case are depicted in Figure~\ref{fig:dNdS} by
the solid blue lines, which are shown only above the estimated
sensitivity of the analysis (see below).  On the contrary, to have a
more robust and realistic estimate of the uncertainty bands we keep
the band resulting from the fits with multiple breaks (for the three
energy bands below 10\,GeV).  These bands are plotted as blue shaded
regions in the figure.  The resulting \dnds\ distributions are
compared to counts of cataloged point sources\footnote{The method of
  deriving \dnds\ for cataloged sources is explained in \pzc,
  Section~4.3.5.}, derived from the 3FGL \citep{2015ApJS..218...23A},
1FHL \citep{2013ApJS..209...34A}, and 2FHL \citep{2016ApJS..222....5A}
source catalogs, respectively.

As demonstrated in the figure, the \opdf\ fits match the
\dnds\ distributions of cataloged sources well within uncertainties.
The \opdf\ method allows us to measure the energy-dependent \dnds\ in
the regime of undetected faint point sources down to integral fluxes
of $\sim\!10^{-11}\,\mathrm{cm}^{-2}\,\mathrm{s}^{-1}$, which are
typically an order of magnitude below the nominal catalog detection
threshold, below which the catalog detection efficiency is much less
than $1$.  The uncertainty bands of the fits significantly broaden
below the sensitivity limit of
$\sim\!10^{-11}\,\mathrm{cm}^{-2}\,\mathrm{s}^{-1}$.  The sensitivity
can be compared to an analytic estimate $S_\mathrm{sens}$,
corresponding to two photons per pixel, indicated by the dashed
vertical lines in Figure~\ref{fig:dNdS} (see \pzc, Section~4). The
actual sensitivity matches these expectations. In the 50\,GeV to
171\,GeV band, the actual sensitivity is better by a factor of
2\,-\,3.

We conclude that the \dnds\ distributions in the four bands below
50\,GeV are compatible with broken power laws for fluxes above
\mbox{$\sim\,10^{-11}\,\mathrm{cm}^{-2}\,\mathrm{s}^{-1}$}.  The
power-law index $n_2$ below the break is compatible with values
between 1.95 and 2.0\,. The \dnds\ distribution in the highest-energy
band between 50\,GeV and 171\,GeV is compatible with a simple power
law\footnote{That is, the fit did not prefer a break significantly
  above the sensitivity limit.} with an index $2.2^{+0.7}_{-0.3}$.
Within uncertainties, this index is compatible with the 2FHL catalog
\dnds\ of \cite{2015arXiv151100693T}, who conducted a catalog analysis
of \mbox{\texttt{Pass 8}} data between 50\,GeV and 2\,TeV.

Given the absence of a significant second (first, for the last band)
intrinsic break of \dnds, we derived corresponding upper limits
(cf.~\pzc). In this case, we assumed that a break would be present if
the indices of the power-law components above and below the break
differed by $\Delta n_{i\,i+1} = |n_i - n_{i+1}| > 0.3$, for
$i=2\,(1)$. The upper limits at the 95\% confidence level are quoted
in Table~\ref{tab:dnds_comp}. All upper limits are either located at
or below the detection thresholds of current catalogs. For the
highest-energy band between 50\,GeV and 171\,GeV, any break has been
constrained to be at fluxes below $1.3\times
10^{-11}\,\mathrm{cm}^{-2}\,\mathrm{s}^{-1}$.  This upper limit is
consistent with the break found in \cite{2015arXiv151100693T}.

\subsection{Anisotropies}\label{ssec:res_aniso}
The anisotropy (or auto-correlation angular power spectrum) provides a
complementary measure of unresolved point sources
\citep[][]{Ackermann:2012uf,
  Cuoco:2012yf,DiMauro:2014wha,2014JCAP...01..049R}.  For a given
\dnds, the anisotropy can be calculated \citep{Cuoco:2012yf} as
\begin{equation}\label{eq:Cp}
C_\mathrm{P}(S_{\rm th}) =  \int_0^{S_{\rm th}} S^2
\frac{\mathrm{d}N}{\mathrm{d}S} \mathrm{d}S,
\end{equation}
where $S_{\rm th}$ is the flux threshold of individually resolved
(detected) point sources. In our case, the integral is effectively
limited to the interval $[S_\mathrm{nd},S_\mathrm{th}]$, given that
the \dnds\ distribution was parameterized with a sharp cutoff below
the node $S_\mathrm{nd}$.  Table~\ref{tab:dnds_comp} lists the
resulting anisotropies corresponding to our \dnds\ fits, assuming flux
thresholds approximating the catalog detection thresholds (quoted in
the caption of the table), with the aim of comparing them with new
anisotropy measurements available in the near future
\citep{Fornasa2016}.

\subsection{Composition of the Gamma-ray Sky}\label{ssec:res_comp}
The contribution of point sources to the total flux $F_\mathrm{tot}$
of the region of interest (ROI) in the energy band
$[E_\mathrm{min},E_\mathrm{max}]$ is given by
\begin{equation}\label{eq:F_ps}
F_\mathrm{ps} = \int_0^{S_{\rm cut}} S \frac{\mathrm{d}N}{\mathrm{d}S} \mathrm{d}S,
\end{equation}
which again is effectively limited to a lower bound of $S_\mathrm{nd}$
in our case.  The full posterior was employed to derive the profile
likelihood of $F_\mathrm{ps}$.  The GF contribution
$F_\mathrm{gal}$ was obtained from the integral template flux and the
normalization parameter $A_\mathrm{gal}$. 
The diffuse isotropic background component $F_\mathrm{iso}$
was already one of the fit parameters.  The sum of the three
components can be compared to $F_\mathrm{tot}$, which was
independently derived from integrating the events map 
divided by the energy-averaged exposure map over the ROI.

Table~\ref{tab:dnds_comp} lists the composition of the high-latitude
gamma-ray sky for each energy band. The contribution of point sources
$F_\mathrm{ps}$ can be compared to the extragalactic gamma-ray
background (EGB), $F_\mathrm{EGB}$, as measured in
\citet{2015ApJ...799...86A}. The resulting fractional contributions
$F_\mathrm{ps}/F_\mathrm{EGB}$ in each energy band are
$0.83^{+0.07}_{-0.13}$, $0.79^{+0.04}_{-0.16}$,
$0.66^{+0.20}_{-0.07}$, $0.66^{+0.28}_{-0.05}$, and
$0.81^{+0.52}_{-0.19}$, respectively.\footnote{The values assume
  reference model A of \citet{2015ApJ...799...86A}.}

\subsection{Comparison with Models}\label{ssec:blazar_models}
In order to assess the power of this method, we compare our
\dnds\ measurement as a function of energy with state-of-the-art
models. We consider all source classes known to provide major
contributions to the EGB.  The blazar gamma-ray luminosity function
(GLF) and spectrum are modeled following \citet{2012ApJ...751..108A}
for FSRQs, \citet{2014ApJ...780...73A} and \citet{2014ApJ...786..129D}
for BL~Lacs, and \citet{2015ApJ...800L..27A} when considering a single
description for all blazars.  Misaligned AGN are taken from
\citet{2014ApJ...780..161D}, while for SFGs we followed
\citet{2012ApJ...755..164A} with an infrared luminosity function from
\citet{2013MNRAS.432...23G}.  The absorption due to extragalactic
background light is modeled according to \citet{2010ApJ...712..238F},
affecting the two highest-energy bands.

\begin{figure}[t]
\epsscale{1}
\plotone{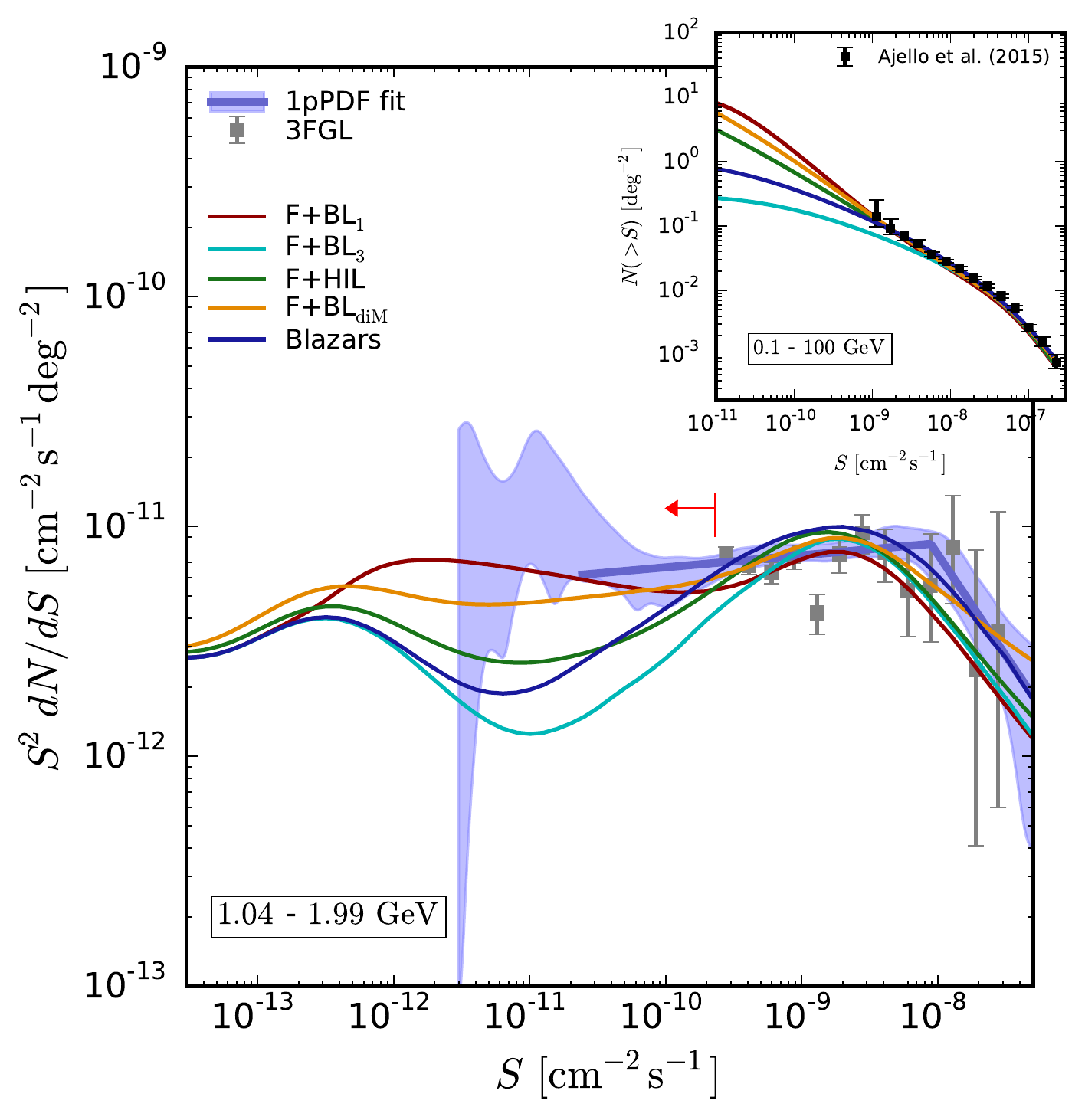}
\caption{Differential source-count distribution \dnds\ in the
  1.04$-$1.99 GeV band, compared to blazar models. The three models
  F+BL$_1$ (solid red line), F+BL$_3$ (solid cyan line), and F+HIL
  (solid green line) include the sum of FSRQs
  \citep{2012ApJ...751..108A} and BL~Lacs, which have been modeled
  assuming the luminosity functions LDDE$_1$, LDDE$_3$, and
  LDDE$_\mathrm{HSP}$ + LDDE$_\mathrm{ISP+LSP}$ of
  \citet{2014ApJ...780...73A}, respectively. The solid orange line
  shows the sum of the FSRQ component from \citet{2012ApJ...751..108A}
  and the BL~Lac model from \citet{2014ApJ...786..129D}. The model
  depicted by the solid dark blue line refers to a single description
  of all blazars by \citet{2015ApJ...800L..27A}. A contribution from
  mAGN \citep{2014ApJ...780..161D} has been added to each model. The
  vertical red line indicates the upper limit (95\% CL) on a second
  intrinsic break.  The inset compares the integral source-count
  distribution $N(>S)$ between 0.1\,GeV and 100\,GeV
  \citep[see][]{2015ApJ...800L..27A} to the models (not including
  mAGN). \label{fig:models}}
\end{figure}

The inset of Figure~\ref{fig:models} shows that all the different
descriptions adopted for blazars are compatible with the cataloged
integral source-count distribution. However, they can have
significantly different behaviors for unresolved sources.

The sum of the model predictions for FSRQs and BL~Lacs \citep[model
  LDDE$_1$ in][]{2014ApJ...780...73A} can reproduce the
\opdf\ measurement fairly well in all five energy bands, as
demonstrated in Figure~\ref{fig:dNdS}.  Misaligned AGN enter the
overall \dnds\ distribution below the threshold of the current
analysis and therefore cannot be constrained (see
Figure~\ref{fig:dNdS}).  Due to their intrinsic faintness, SFGs start
to contribute only at very low, unconstrained fluxes.

However, Figure~\ref{fig:models} shows that deviations in the faint
end predicted by the other blazar models may be in tension with the
allowed region derived from the \opdf\ analysis. A comprehensive study
of the implications for blazar models is beyond the scope of this
Letter, but we can conclude that the methodology of using the
\opdf\ for measuring the gamma-ray source-count distribution has
excellent sensitivity for probing unresolved blazars and the faint
part of the blazar GLF.

\acknowledgments
We acknowledge the valuable support by the internal referee Marco
Ajello and the anonymous journal referee in improving the manuscript,
and the support of the {\sl Servizio Calcolo e Reti} of the Istituto
Nazionale di Fisica Nucleare, Sezione di Torino.

This work is supported by the research grant {\sl Theoretical
  Astroparticle Physics} number 2012CPPYP7 under the program PRIN 2012
funded by the Ministero dell'Istruzione, Universit\`a e della Ricerca
(MIUR), by the research grants {\sl TAsP (Theoretical Astroparticle
  Physics)} and {\sl Fermi} funded by the Istituto Nazionale di Fisica
Nucleare (INFN), and by the {\sl Strategic Research Grant: Origin and
  Detection of Galactic and Extragalactic Cosmic Rays} as well as {\sl
  Excellent Young PI Grant: The Particle Dark-matter Quest in the
  Extragalactic Sky} funded by Torino University and Compagnia di San
Paolo.

The \textit{Fermi}-LAT Collaboration acknowledges support for LAT
development, operation and data analysis from NASA and DOE (United
States), CEA/Irfu and IN2P3/CNRS (France), ASI and INFN (Italy), MEXT,
KEK, and JAXA (Japan), and the K.A.~Wallenberg Foundation, the Swedish
Research Council and the National Space Board (Sweden). Science
analysis support in the operations phase from INAF (Italy) and CNES
(France) is also gratefully acknowledged.

\bibliographystyle{apj}

\end{document}